\documentclass{aa}  

\usepackage{graphicx}
\usepackage{txfonts}

\usepackage[
        colorlinks=true,
        linkcolor=blue,
        citecolor=blue,
        filecolor=black,
        urlcolor=red,
        bookmarks=true,
        bookmarksopen=true,
        bookmarksopenlevel=3,
        plainpages=false,
        pdfpagelabels=true,
        breaklinks=true,
]{hyperref}
\usepackage{wasysym}
\usepackage[english=american]{csquotes}
\usepackage{longtable}
\usepackage{booktabs}
\usepackage{subfigure}
\usepackage[multiple,symbol]{footmisc}
\bibpunct{(}{)}{;}{a}{}{,}

\begin{document}

   \title{Non-detection of TiO and VO in the atmosphere of WASP-121b using high-resolution spectroscopy}
   \titlerunning{Non-detection of TiO and VO in WASP-121b}

   \author{S. R. Merritt
          \inst{1},
          N. P. Gibson
          \inst{2},
          S. K. Nugroho
          \inst{1},
          E. J. W. de Mooij
          \inst{1},
          M. J. Hooton
          \inst{3, 1},
          S. M. Matthews
          \inst{1},
          L. K. McKemmish
          \inst{4,} \inst{5},
          T. Mikal-Evans
          \inst{6},
          N. Nikolov
          \inst{7},
          D. K. Sing
          \inst{7},
          J. J. Spake
          \inst{8, 7}
          \and
          C. A. Watson\inst{1}
          }
    \authorrunning{S. R. Merritt et al.}

   \institute{Astrophysics Research Centre, School of Mathematics and Physics, Queen's University Belfast, Belfast BT7 1NN, UK\\
              \email{smerritt01@qub.ac.uk}
         \and
             School of Physics, Trinity College Dublin, Dublin 2, Ireland
        \and
            Physikalisches Institut, Universität Bern, Gesellschaftsstrasse 6, 3012, Bern, Switzerland
        \and
            School of Chemistry, University of New South Wales, 2052 Sydney, Australia
        \and
            Department of Physics and Astronomy, University College London, Gower Street, WC1E 6BT London, UK
        \and
            Department of Physics, Massachusetts Institute of Technology, Cambridge, MA 02139, USA
        \and
            Department of Earth and Planetary Sciences, Johns Hopkins University, Baltimore, MD, USA
        \and
            Physics and Astronomy, Stocker Road, University of Exeter, Exeter, EX4 3RF, UK
             }

   \date{Received XXX; accepted YYY}

  \abstract
   {Thermal inversions have long been predicted to exist in the atmospheres of ultra-hot Jupiters. However, detection of two species thought to be responsible -- TiO and VO -- remain elusive. We present a search for TiO and VO in the atmosphere of the ultra-hot Jupiter WASP-121b ($T_\textrm{eq} \gtrsim 2400$ K), an exoplanet already known to show water features in its dayside spectrum characteristic of a temperature inversion as well as tentative evidence for VO at low-resolution. We observed its transmission spectrum with UVES/VLT and used the cross-correlation method -- a powerful tool for the unambiguous identification of the presence of atomic and molecular species -- in an effort to detect whether TiO or VO were responsible for the observed temperature inversion. No evidence for the presence of TiO or VO was found at the terminator of WASP-121b. By injecting signals into our data at varying abundance levels, we set rough detection limits of $[\text{VO}] \lesssim  -7.9$ and $[\text{TiO}] \lesssim -9.3$. However, these detection limits are largely degenerate with scattering properties and the position of the cloud deck. Our results may suggest that neither TiO or VO are the main drivers of the thermal inversion in WASP-121b, but until a more accurate line list is developed for VO, we cannot conclusively rule out its presence. Future work will search for finding other strong optically-absorbing species that may be responsible for the excess absorption in the red-optical.}


   \maketitle
%

\section{Introduction}

\label{sec:intro}
An open problem in the field of exoplanet atmospheres is that of the existence and origin of thermal inversions in hot Jupiters. It was postulated by \citet{Hubeny2003} and \citet{Fortney2008} that the hottest of hot Jupiters ($T_\text{eq} > 2000$ K) would host titanium oxide and vanadium oxide in their gaseous form, which would strongly absorb UV and visible light and heat the upper atmosphere, leading to a rise in temperature with altitude: a thermal inversion. However, evidence for the existence of thermal inversions has proven unexpectedly difficult to discover, as have any direct detections of TiO or VO.

The first evidence of an inversion layer was reported by \citet{Knutson2008} in the atmosphere of HD 209458b, using secondary eclipse observations from the Spitzer Space Telescope. However, later work by \citet{Hansen2014} showed that several previous single-transit measurements made using Spitzer had much higher uncertainties than previously thought, and later work by \citet{Zellem2014}, \citet{DiamondLowe2014}, \citet{Schwarz2015}, \citet{Evans2015} and \citet{Line2016} showed results consistent with no temperature inversion layer. Further work found a surprising number of flat, blackbody-like emission spectra for a number of ultra-hot Jupiters (\citealt{Swain2013}; \citealt{Cartier2017}; \citealt{Kreidberg2018}; \citealt{Arcangeli2018}; \citealt{Mansfield2018}) showing none of the emission or absorption features that would indicate the presence of the expected temperature inversion. \citet{Sedaghati2017} reported a detection of TiO in the ground-based transmission spectrum of WASP-19b using the VLT, contradicting an earlier result from HST/STIS that found no evidence of TiO \citep{Huitson2013}. However, a subsequent attempt to reproduce the detection using IMACS on GMT was unsuccessful \citep{Espinoza2019}. \citet{Haynes2015} observed the dayside of WASP-33b ($T_{\text{eq}} \sim 2700$K) at low resolution with WFC3 on HST and found a thermal emission feature at $\sim$\,1\,\,micron, consistent with the presence of TiO. This claim was later strengthened by a detection of TiO in the dayside spectrum of WASP-33b using high-resolution spectroscopy and the cross-correlation technique by \citet{Nugroho2017} with observations taken by HDS on the Subaru telescope, and remains the strongest claim for TiO in the presence of an exoplanet atmosphere.

Most searches for temperature inversions and the species believed to cause them have been performed using low-resolution spectroscopy in transmission and emission, using powerful techniques which are often used to characterise exoplanet atmospheres (\citealt{Seager2000}; \citealt{Brown2001}). Transmission spectroscopy measures the apparent change in radius with wavelength as a planet passes in front of its host star during transit, divining a wealth of information on the thermal structure of the atmosphere at the limb, the atomic and molecular species present within, and the potential presence of clouds and hazes. Emission spectroscopy aims to directly measure the exoplanet's thermal spectrum and provides similar information on the dayside atmosphere of the planet, although it probes different altitudes and utilises different geometry, leading to a differing sensitivity to trace elements.

Due to the need for extremely stable time-series data (to a precision of ${\sim}10^{-4}$), much transmission and emission spectroscopy has been performed using space-based facilities such as the Hubble Space Telescope and Spitzer (e.g. \citealt{Charbonneau2002}; \citealt{Pont2008}; \citealt{Huitson2012}; \citealt{Berta2012}; \citealt{Pont2013}; \citealt{Kreidberg2014}; \citealt{Nikolov2015}; \citealt{Sing2016}). Recently, multi-object spectrographs have provided the necessary stability for ground-based facilities to also play a role (e.g. \citealt{Bean2010}, \citeyear{Bean2011}; \citealt{Crossfield2013}; \citealt{Gibson2013a, Gibson2013b}; \citeyear{Gibson2019}; \citealt{Jordan2013}; \citealt{Stevenson2014}; \citealt{Lendl2016}; \citealt{Mallonn2016}; \citealt{Nikolov2016}, \citeyear{Nikolov2018}). 

The recent adaptation of the high-resolution Doppler-resolved spectroscopy technique (\citealt{Snellen2010}; \citealt{Brogi2012}; \citealt{Birkby2013}), originally used to characterise spectroscopic binary systems, has drastically improved our ability to characterise exoplanet atmospheres from the ground. This technique takes advantage of the large radial velocity of the planet and the corresponding Doppler shift in its atomic and molecular lines. This allows us to isolate the signal of the planet from its parent star and from telluric lines by using detrending techniques to remove static and quasi-static signals from a spectroscopic time-series. This leaves only the Doppler-shifted lines from the planet's transmission spectrum, which (unlike in low-resolution spectroscopy) are individually resolved at high dispersion. Thus, the planetary signal can be extracted from the noise using cross-correlation with model spectra of the predicted transmission spectrum for individual atomic or molecular species, allowing us to effectively sum up over several resolved spectral lines (often numbering in the thousands) and thereby strengthen the detection signal. This method has been used to detect evidence of molecules such as CO, H$_{2}$O, TiO, CH$_4$ and HCN; and atomic species such as Fe, Ti, Mg, Na, K and Ca (\citealt{Snellen2010}; \citealt{Brogi2012}; \citealt{Birkby2013}; \citealt{Lockwood2014}; \citealt{Nugroho2017}; \citealt{Hoeijmakers2018}, \citeyear{Hoeijmakers2019}; \citealt{Cabot2019}; \citealt{Guilluy2019}; \citealt{Turner2019}; \citealt{Yan2019}; \citealt{Gibson2020}). The extra velocity information encoded in this technique has also allowed for the detection of the rotational velocity of the planet \citep{Snellen2014} and planetary winds \citep{Brogi2016}, and greater constraints on the mass of non-transiting planets due to the direct measurement of the planetary radial velocity \citep{Brogi2014}. It is thus ideally suited to work in tandem with low-resolution spectroscopy to confirm the presence of atomic and molecular species.

Here we present UVES observations of WASP-121b \citep{Delrez2016}, an ultra-hot Jupiter which orbits a bright F6V-type star (V = 10.5, \citealt{Hoeg2000}) with a period of just 1.27 days, resulting in an equilibrium temperature of over 2400 K. This places it well within the temperature regime of ultra-hot Jupiters believed to host temperature inversions. The combination of its high temperature and its highly inflated radius also make WASP-121b an excellent subject for transmission and emission spectroscopy. Recent UV observations have found that WASP-121b has an extended and potentially escaping atmosphere, with evidence for \ion{Fe}{ii} and \ion{Mg}{ii} during transit extending far higher in the atmosphere than previously-detected features at redder wavelengths \citep{Sing2019}, and high-resolution spectroscopic analysis using UVES \citep{Gibson2020} and HARPS \citep{Bourrier2020, Cabot2020} has also discovered the presence of \ion{Fe}{i} deeper in the atmosphere. An excess in UV absorption was previously reported by \citet{Salz2019}, and eclipse observations in the $z^{\prime}$ band have determined upper limits on the planet's albedo \citep{Mallonn2019}.

WASP-121b has also proven to be a rich target in the search for temperature inversions. Low-resolution observations with HST have provided a detection of water and tentative evidence for TiO or VO in transmission (\citealt{Evans2016}; \citealt{Tsiaras2018}); more evidence for VO and the first direct measurement of a temperature inversion via water features detected in emission \citep{Evans2017}; evidence for VO (although not TiO) and an unknown blue absorber suggested to be SH in transmission \citep{Evans2018}; additional emission measurements and a new retrieval methodology also suggesting the presence of VO \citep{Mikal-Evans2019}; and phase-curve photometry from TESS which confirms the temperature inversion and suggests TiO, VO or H$^{-}$ as the cause (\citealt{Daylan2019}; \citealt{Bourrier2019}). This wealth of information concerning the existence of TiO or VO at low-resolution, in combination with excellent evidence for the presence of a temperature inversion due to water emission features, makes WASP-121b an excellent subject for study at high-resolution in order to confirm the identity of the species causing the observed temperature inversion.

In the following pages we present the results of a search for TiO and VO in the atmosphere of the ultra-hot Jupiter WASP-121b using high-resolution spectra from the UV-Visual Echelle Spectrograph (UVES) at the Very Large Telescope (VLT) \citep{Dekker2000}, an instrument which has previously proven successful in characterising exoplanet atmospheres (\citealt{Snellen2004}; \citealt{Czesla2015}; \citealt{Khalafinejad2017}; \citealt{Gibson2019}). A first look at the blue arm of these observations was presented in \citet{Gibson2020}. In Sec.~\ref{sec:obs} we describe the observations and the extraction of the data. In Sec.~\ref{sec:anal}, we discuss the data processing steps, the atmospheric models used, and the search for TiO and VO using the cross-correlation method. Section~\ref{sec:res} details our non-detection and the resulting injection tests used to set detection limits for TiO and VO. Sections~\ref{sec:discus} and~\ref{sec:con} discuss the results and their implications.

\section{Observations}
\label{sec:obs}

\begin{figure}
	\includegraphics[width=\columnwidth]{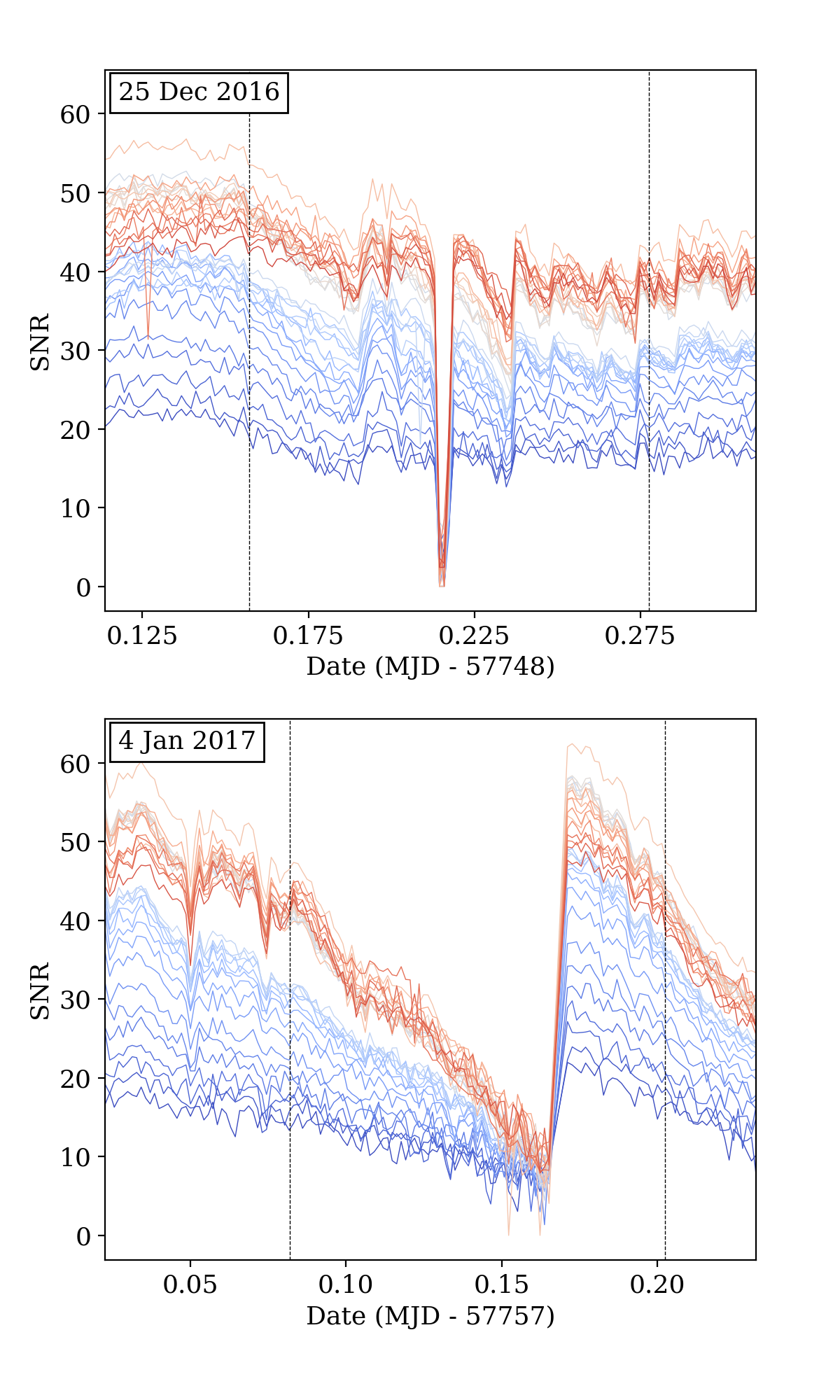}
    \caption{Peak signal-to-noise for a single pixel element over time for the first night of observations (top) and the second (bottom), showing all echelle orders. The colour scale represents the wavelength range of the echelle orders from blue to red. The times of ingress and egress are indicated by the black dotted lines. The strong dip in SNR on the first night was caused by loss of guiding: the three exposures affected were removed from our analysis. The SNR of the second night was affected by pointing drift and variable seeing.}
    \label{fig:SNR}
\end{figure}

Two transits of WASP-121b were observed on the nights of December 25 2016 and January 4 2017 as part of program 098.C-0547 (PI: Gibson) using the high resolution echelle spectrograph UVES, mounted on the 8.2m "Kueyen" (UT2) telescope of the VLT. A "free template" with dichroic \#2 and cross-dispersers \#2 and \#4 was utilised, and image slicer \#3 was used to maximise throughput and spectral resolution with a slit length of 10.0 arcsec. This resulted in R ${\sim}~$110,000 from 565 nm to 946 nm over a range of 33 spectral orders in the red arm, with a central wavelength of 760 nm. For this paper we focus on analysis of the red arm data only. Analysis and a search for features in the blue arm can be found in \citet{Gibson2020}. 

An exposure time of 100 seconds and a read-out time of ${\sim}24$ seconds was used for both nights, resulting in a cadence of 124 seconds. The first transit observation covered 4.7 hours in 137 exposures: 28 before ingress, 83 in-transit and 26 after egress. Guiding was lost for exposures 70-72, and these were subsequently removed from our analysis. The second transit observations covered 5.0 hours and resulted in 143 exposures, of which 39 were pre-ingress, 80 in-transit and 24 post-egress. The small difference in the number of in-transit observations between the two nights was caused by a short pause in observations during the first night in order to correct the guiding.

Data were analysed using a custom pipeline written in \texttt{Python}, similar to but independent from the pipeline outlined in \citet{Gibson2019}, against which it was benchmarked. This pipeline performed basic calibrations and extraction of the time-series spectra for each spectral order. The UVES \texttt{esorex} pipeline (version 5.7.0) was used to generate flat-field and bias frames, provide an initial wavelength solution and determine the order trace positions. A simple aperture extraction was performed with an aperture width of 55 pixels, chosen to comfortably contain all five spectral traces for each order while preventing any potential overlap. Optimal extraction had been tested previously in the similar pipeline outlined in \citet{Gibson2019}, and was found to make little difference to the SNR of the extracted spectra, as expected for bright targets where it can be assumed that background and read noise are negligible compared to the stellar spectrum. As the five spectral traces almost entirely fill the length of the extraction aperture, no background subtraction was performed, as the difference was found to be negligible on the same data. 

Our custom pipeline allowed finer control over both the final data products and the aperture size and centering during the extraction process. We chose to extract and use the raw flux from each spectral order, rather than merging the orders after performing resampling, flat-fielding and blaze correction. This was preferred as the instrumental response changes in time and cannot be corrected by flat-fielding, and also ensures optimal preservation of spectral information at the order edges. The stability of the wavelength solution was checked by cross-correlating the spectral time-series for each order by a median spectrum of that order: total drift over each night was found to be $\sim$1.3 $\text{km s}^{-1}$, significantly smaller than a resolution element ($\sim$2.5 $\text{km s}^{-1}$ at R $\sim$ 110,000) and thus the original wavelength solution was deemed to be fit for our purposes.

The throughput of the UVES red arm peaks at the red end of the wavelength range, with a typical signal-to-noise of 30 to 45 per spectral element in the centre of the orders for the first night. Due to issues with both seeing and pointing, throughput for the second night was far more variable, with SNR ranging from 2 to 55 and the lowest count-rates occurring predominantly during the transit. We therefore discarded the second transit from our analysis and present analysis from the first night only. The peak SNR for each order over the course of both nights is presented in Fig.~\ref{fig:SNR}: the time-dependent behaviour shown was largely consistent across all pixel elements for a given order. The barycentric Julian date (BJD$_{\text{TDB}}$) was calculated for each observation using \texttt{astropy.time} routines.

Outliers were removed from each order using a two-stage process. A two-component PCA reconstruction was subtracted from the data to remove all time- and wavelength-stable trends. Five iterations of sigma-clipping were performed upon the resulting residuals to mask all outliers above $3\sigma$: their values were then set to zero. The PCA reconstruction was then added back onto the data, replacing any clipped outliers with their value from the PCA reconstruction. A few very strong outliers still remained in a small number of orders, and these were removed by performing a further round of sigma-clipping upon the raw spectra and replacing the values with the median of the surrounding 100 pixels. The cleaning process replaced an average of 0.6\% of pixels per order, and is not expected to affect our analysis.

\section{Analysis}
\label{sec:anal} 

\subsection{Pre-processing steps}
\label{sub:prepro}
\begin{figure*}
	\includegraphics[width=\textwidth]{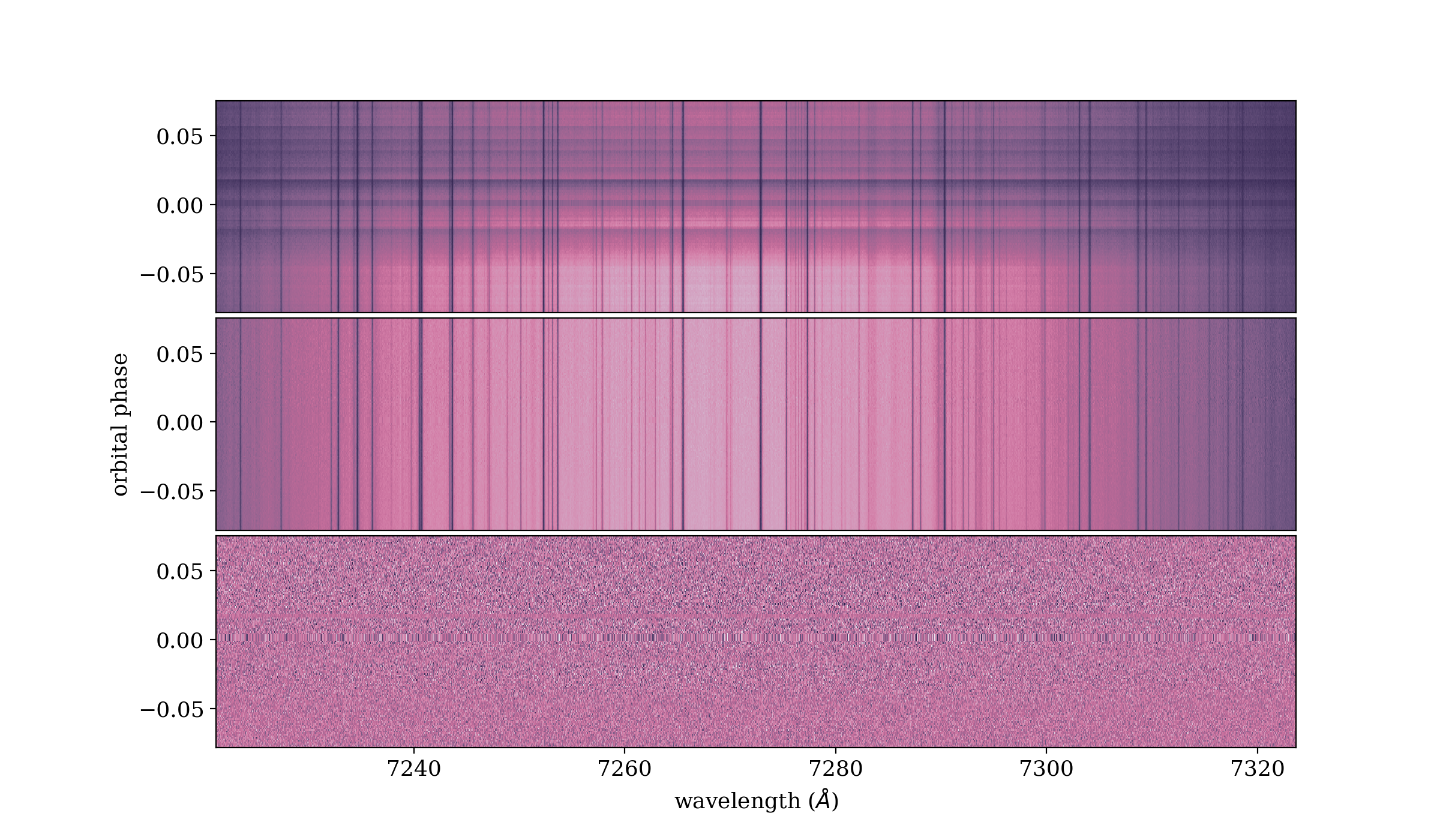}
    \caption{Data-processing steps shown for a single echelle order. Top: the raw extracted time-series spectra after cleaning of outliers. Middle: the time-series spectra after blaze-correction. Bottom: detrended residuals from the division of the \textsc{SysRem} model weighted by the uncertainties. Minor structure remains in the residuals due to lower weighting of low-SNR orders caused by poorer seeing (horizontal banding) and due to imperfect removal of time-dependent tellurics (vertical banding at bottom). The telluric structure does not persist into the in-transit frames and is therefore weighted-out in later analysis.}
    \label{fig:meth}
\end{figure*}

We follow a methodology now common to searches for molecular and atomic species in high-resolution spectral time series (\citealt{Snellen2010}; \citealt{Brogi2012}; \citealt{Birkby2013}; \citealt{Nugroho2017}). The first step is to remove all trends within the data that are static or quasi-static in time, leaving behind -- in theory -- the Doppler-shifted planetary signal (after removal of its continuum) plus photon noise. An example of our data processing on a single order is shown in Fig.~\ref{fig:meth}. 

First, the variation in the blaze function is removed. This is accomplished by dividing each order of time-series spectra by the median spectrum, then smoothing the resulting spectral residuals with a median filter with a width of 15 pixels and a Gaussian filter with a standard deviation of 50 pixels, creating a smoothed map of the blaze distortion per order. The original time-series spectra are then divided through by the blaze distortion map. Instead of removing the blaze function altogether, this places each spectrum on a "common" blaze. This is done as any common trends in time will be removed later in the pre-processing, and ensures spectral information is optimally preserved.  The blaze correction was found to be unstable at the blue end of each spectral order, where the signal-to-noise is very low: subsequently, the first 500 pixels of each order were removed from our analysis. This removed area represents only 12\% of each order of 4096 pixels and lies entirely within the overlapping wavelength region of the echelle orders, so its removal is not expected to be significant.

Next, the quasi-static telluric and stellar spectral features are removed from the time-series using the \textsc{SysRem} algorithm \citep{Tamuz2005}. First used in this context by \citet{Birkby2013}, \textsc{SysRem} has become a standard tool for the detrending of high-resolution time-series spectra (e.g. \citealt{Birkby2017}, \citealt{Nugroho2017}). The main advantage \textsc{SysRem} retains over other PCA-based detrending methods lies in its inherent treatment of the pixel uncertainties. Rather than simply using Poisson-noise statistics, we estimate the pixel uncertainties by taking an outer product of the time-wise and wavelength-wise variance of each order. This was done due to the removal of the stellar spectrum inherent to the cross-correlation technique. Poisson-noise estimates based on count-rates and read-noise will impose lower uncertainties on negative values of the residuals, causing a bias in the uncertainty determination \citep{Gibson2020}. 

We use \textsc{SysRem} to create 2D model representations of each time-series order as an outer product of two column vectors. This model is then subtracted from the data and the next \textsc{SysRem} iteration determined from the residuals, which is subtracted again. However, instead of using the resulting iteratively model-subtracted residuals, we instead sum the \textsc{SysRem} models for each iteration and divide each time-series order by this model. This has the effect of preserving the relative line-strengths of the planet's transmission spectrum. The detrended residuals are then weighted by the pixel uncertainties, which ensures that noisier, low-flux areas of the time-series contribute less to the analysis.

Finally, the spectra are super-sampled to 2/3 of the average pixel width for each order and corrected to the stellar rest frame in a single linear interpolation, adjusting for both the barycentric velocity variations and the systemic velocity, and the wavelength scale is linearised in $\Delta \lambda$.

\subsection{Model transmission spectra}
\label{sub:model}
Constraining the existence of molecular or atomic features using the high-resolution cross-correlation technique requires models of the planet's atmosphere to act as cross-correlation templates. These models are generated using high-temperature line-lists for the species of interest to generate absorption cross-sections which can be fed into radiative transfer models to create template spectra for the cross-correlation process. However, the cross-correlation technique is highly sensitive to line position, and small inaccuracies in line lists can result in the reduction or non-detection of exoplanetary signals \citep{Hoeijmakers2015}.

\begin{table}
\centering
\begin{tabular}{cc}
Parameter						& Value 								\\
\hline \hline
																\\
\textbf{WASP-121} 				& 									\\
\hline
																\\
$M_{\star}~(M_{\odot})$			& $1.353^{+0.080}_{-0.079}{~}^\text{a}$			\\
$R_{\star}~(R_{\odot})$			& $1.458 \pm 0.080{~}^\text{a}$				\\
Spectral type 					& F6V${~}^\text{a}$ 						\\
T$_{\textrm{eff}}$ (K) 			& $6459 \pm 140{~}^\text{a}$					\\
$V$-magnitude					& $10.44{~}^\text{a}$ 						\\
$v_{\textrm{sys}}$ ($\text{km s}^{-1}$) 				& $38.36 \pm 0.43{~}^\text{b}$					\\
							&									\\
\textbf{WASP-121b} 				&									\\
\hline
																\\
$T_{0}~(\textrm{BJD}_{(TDB)})$ 	& $2457599.551478 \pm 0.000049{~}^\text{c}$	\\
$P$ (days)					& $1.2749247646 \pm 0.0000000714{~}^\text{c}$	\\
$a/R_{\star}$ 					& $3.86  \pm 0.02{~}^\text{d}$					\\
$R_{p}/R_{\star}$				& $0.1218 \pm 0.0004{~}^\text{d}$				\\
$i~(\deg)$						& $89.1 \pm 0.5{~}^\text{d}$					\\
$T_{\textrm{eq}}$ (K)			& $> 2400$							\\
$H$ (km)						& $\sim 550^\text{d}$							\\
$K_p$ ($\text{km s}^{-1}$)                    & $\sim 217 $ \\
\\
\end{tabular}
\caption{Table of stellar and planetary parameters for the WASP-121b system utilised in this paper. Values marked with (a) are adopted from \citet{Delrez2016}; (b) from \citet{Gaia2018}; (c) from \citet{Sing2019}; and (d) from \citet{Evans2018}.}
\label{table:param}
\end{table}

Model spectra were generated for TiO and VO using a method similar to that outlined in Nugroho et al. (\textit{submitted}),  with planetary and atmospheric parameters taken from \citet{Evans2018} in order to provide the most direct comparison to their transmission spectrum. The full list of system and planetary parameters used in this work is presented in Table~\ref{table:param}. 

To calculate the cross-section of VO, the \textsc{ExoMol} line list (\citealt{McKemmish2016}, \citealt{TennysonYurchenko2012}) was used. The TiO cross-sections were generated using three different line lists: \citet{Plez1998} in both its original 1998 incarnation and the updated, corrected 2012 version, and the new \textsc{ToTo} line list from \textsc{ExoMol} \citep{McKemmish2019}. This was done to allow us to benchmark the performance of the different TiO line lists against each other, both in the event of a detection and in our injection tests, as line lists are known to vary considerably. The Plez 2012 and \textsc{ToTo} line lists both explicitly include experimental data: Plez 2012 incorporates experimental line lists and \textsc{ToTo} includes experimentally-derived energy levels. Conversely, the VO line list and Plez 98 are derived entirely from quantum chemistry calculations and are thus thought to be less accurate and potentially inappropriate for high-resolution cross-correlation. We discuss this issue further in Sec.~\ref{sec:discus}.

For VO, we only considered $^{51}$V$^{16}$O. For TiO, we considered the five most stable isotopologues: $^{46}$Ti$^{16}$O, $^{47}$Ti$^{16}$O, $^{48}$Ti$^{16}$O, $^{49}$Ti$^{16}$O, $^{50}$Ti$^{16}$O. The cross-section was calculated using \texttt{HELIOS-K} \citep{GrimmHeng2015} assuming a Voigt line profile with thermal and natural broadening only at a resolution of 0.01 cm$^{-1}$ with an absolute line wing cut-off of 100 cm$^{-1}$.

\begin{figure*}
	\includegraphics[width=\textwidth]{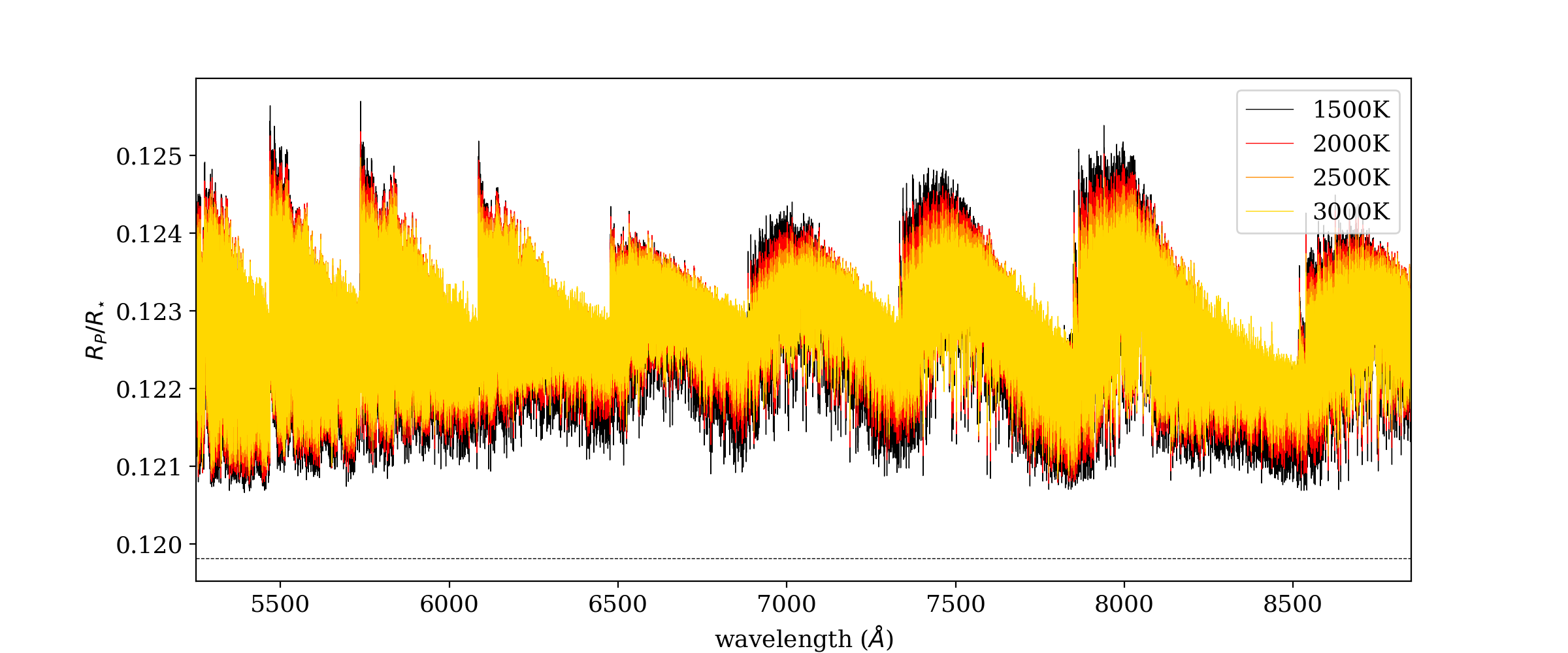}
    \caption{Example model spectrum of TiO with VMR = 10$^{-6}$ before continuum removal, with all four temperatures overplotted to show the change in relative line strength with temperature. The black dotted line indicates the position of the cloud deck.}
    \label{fig:model}
\end{figure*}

We produced the transmission spectrum of WASP-121b by assuming a 1D plane-parallel isothermal atmosphere divided into a hundred layers evenly spaced in log pressure from 2 to 10$^{-12}$ bar. All spectra were generated at four different temperatures of 1500 K (consistent with \citet{Evans2018}), 2000 K, 2500 K and 3000 K in order to cover the range of potential temperatures at the limb. The scale height was kept fixed at 550 km to provide the best match to the low-resolution transmission spectrum measured by \citet{Evans2018}. An example of one of the spectra and its change with temperature is shown in Fig.~\ref{fig:model}.

For both elements, we assumed a constant volume mixing ratio (hereafter VMR) at all pressures, ranging from 10$^{-13}$ to 10$^{-3}$. For VO we also assumed a VMR of 10$^{-6.6}$, consistent with the detection of \citet{Evans2018}. We included Rayleigh scattering by H$_{2}$ and bound-free continuum absorption by H$^{-}$ with a VMR of 5\,$\times$\,10$^{-10}$, and set a cloud deck at P = 20 mbar, again to be consistent with \citet{Evans2018}. The radius of the planet as a function of wavelength was then calculated by integrating the opacity through transit chords.

Lastly, we removed the continuum and the scattering profile from the model spectra by subtracting a forward-backward Butterworth low-pass filter to remove any low-frequency structure from the model spectra caused by large areas of densely overlapping lines, as all such structure will be removed from our spectra by the pre-processing steps outlined in Sec.~\ref{sub:prepro}. These low-frequency structures in the models also tend to "catch" on any remaining structure in our spectra that has been imperfectly removed by \textsc{SysRem}, causing spurious artifacts in the cross-correlation.

\subsection{Cross-correlation}
\label{sub:cc}
As described in Sec.~\ref{sub:prepro}, we divide the orders of our spectra through by a \textsc{SysRem} model consisting of 9 iterations to remove the stellar and telluric features. The optimal number of \textsc{SysRem} iterations was arrived at empirically through injection tests, where the aim was to find the lowest number of \textsc{SysRem} iterations that would still provide a strong detection of an injected signal. The number of iterations is minimised as, at high numbers of iterations, \textsc{SysRem} has been found to reduce or remove the exoplanet spectrum \citep{Cabot2019}. Unlike many recent papers utilising the cross-correlation method (e.g. \citealt{Yan2019}), we do not need to model and remove the stellar spectrum or the Rossiter-McLaughlin effect, as TiO and VO are not expected to exist in the atmosphere of WASP-121, an F6V star. As also mentioned in Sec.~\ref{sub:prepro}, we further weight the \textsc{SysRem} residuals by the pixel variance to ensure that noisier areas of the time-series contribute less to the analysis.

After wavelength linearisation and correction to the stellar rest frame, each frame of each order is then cross-correlated with the corresponding section of the model spectra (also at rest), using the \texttt{NumPy} function \texttt{numpy.correlate}. As no planetary signal is expected in the frames occurring out of transit, and any signal will be weaker during ingress and egress, the resulting time-series of cross-correlation functions is then weighted by a simple transit model for WASP-121b using the equations of \citet{MandelAgol2002} and assuming no limb-darkening. 

The cross-correlation function is expected to peak where there is a match between the model template and the spectra; this peak, if extant, will be Doppler-shifted over time, resulting in a slanted trace in the cross-correlation time-series. For very strong molecular signals, this trace can be seen with the naked eye (e.g. \citealt{Snellen2010}). For weaker signals, e.g. those where the molecules are predicted to be less abundant in the atmosphere, the next step is to integrate the cross-correlation time series over a range of predicted planetary radial velocity semi-amplitudes, where the RV at each frame $v_{\text{p}}$ is given by:
\[
v_{\text{p}}(\phi) = v_{\text{sys}} + K_\textrm{p}\sin{(2\pi\phi)},
\]
\noindent where $v_{\text{sys}}$ is the systemic velocity, $K_\textrm{p}$ is the radial velocity semi-amplitude of the planet and $\phi$ is the orbital phase of the planet, where $\phi$ = 0 represents the mid-transit time. Note that here, $v_{\text{sys}}$  is expected to be zero, as the spectra have already been shifted to the stellar rest frame by correcting for both the systemic and the barycentric velocity. The x-axis of the cross-correlation time series can thus be thought of in terms of velocity "lag" from the stellar rest frame. 

We integrate the resulting cross-correlation function time-series for each order by interpolating each order time-series to a predicted radial velocity curve using values for $K_\textrm{p}$ from -500 to +500 $\text{km s}^{-1}$ in steps of 1 km s$^{-1}$, smaller than the average single resolution element of the original spectra ($\sim$ 2.5 $\text{km s}^{-1}$). These are then summed column-wise in time and stacked to create a series of cross-correlation velocity heat maps, one for each order, where the y-axis is now $K_\textrm{p}$ and the x-axis, as previously, is in velocity "lag" from the stellar rest frame, or $v_{\text{sys}}$. These individual order maps are then summed together to form a final map. The large range in $K_\textrm{p}$ was chosen in order to create large cross-correlation maps so that the overall noise profile could be studied. Due to the use of the cross-correlation, the z-axis is in arbitrary units: in order to set a detection significance, the maps are divided through by their standard deviation away from the location of the predicted signal. Our detection threshold is set at 4$\sigma$ due to both our simplistic significance calculation and to the presence of noise fluctuations in the cross-correlation map which can reach this level. As demonstrated in \citet{Cabot2019}, these fluctuations can be mistaken for detections if they appear in the expected position.

This process was repeated for every template described in Sec.~\ref{sub:model}, and a cross-correlation map generated for each template.

\subsection{Results and injection tests}
\label{sec:res}

\begin{figure*}
	\includegraphics[width=\textwidth]{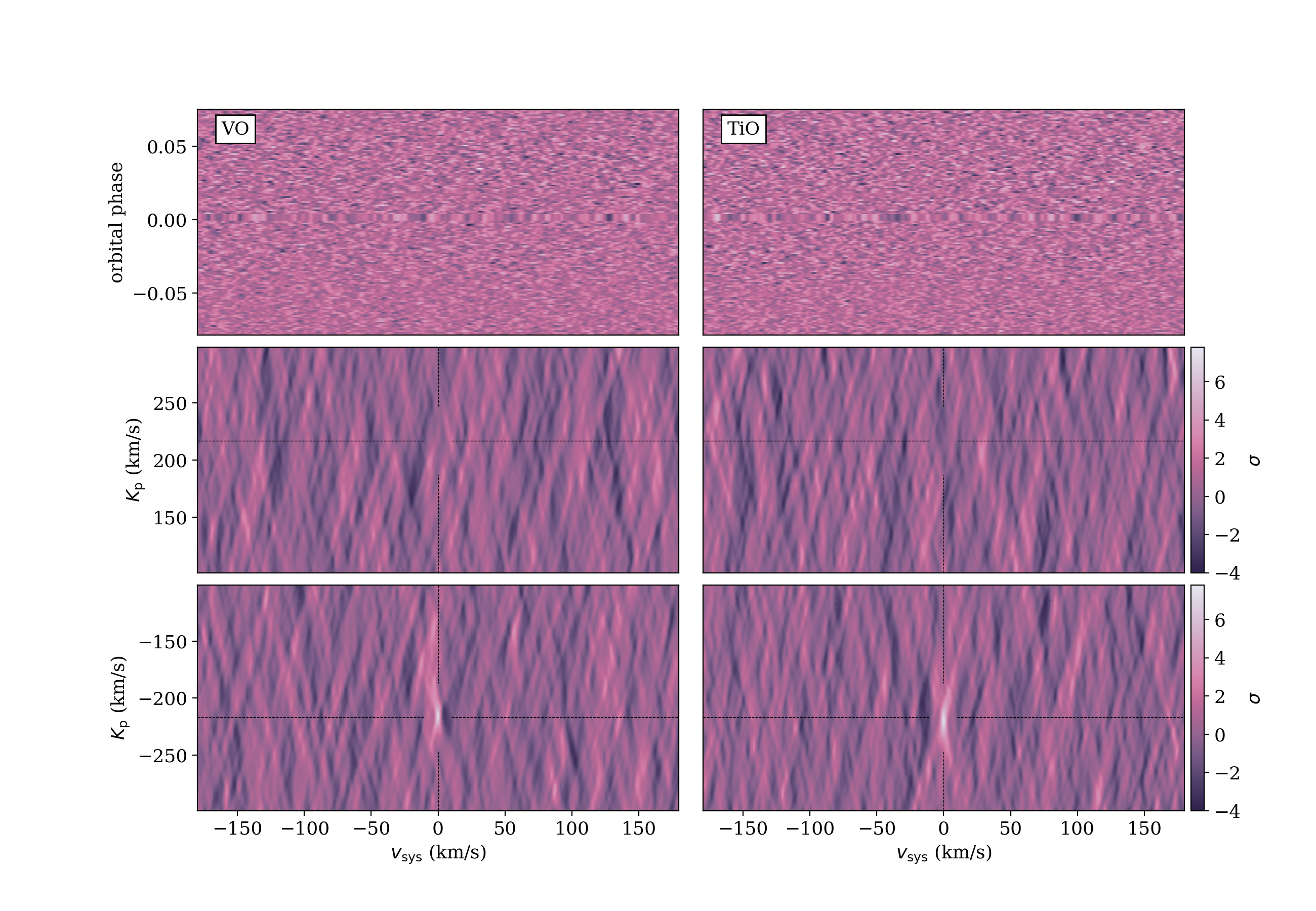}
    \caption{Results from cross-correlation with VO (left) and TiO (right) templates, at a log VMR of -6.6 and -7 respectively, for 1500 K. Top row: cross-correlation function time series summed over all echelle orders, showing no visible trace. Middle: cross-correlation maps summed over all echelle orders, showing non-detections for both species: the black dotted lines indicate the expected velocity position for the signal ($K_\textrm{p}$ = 217 $\text{km s}^{-1}$). Bottom: results from injection tests, showing a detection of the injected signal at 6.9$\sigma$ for VO and 7.2$\sigma$ for TiO at the expected velocity position. Note that the expected value for $v_\textrm{sys}$ is zero as the spectra have been corrected to the stellar rest frame.}
    \label{fig:res}
\end{figure*}

\begin{figure*}
	\includegraphics[width=\textwidth]{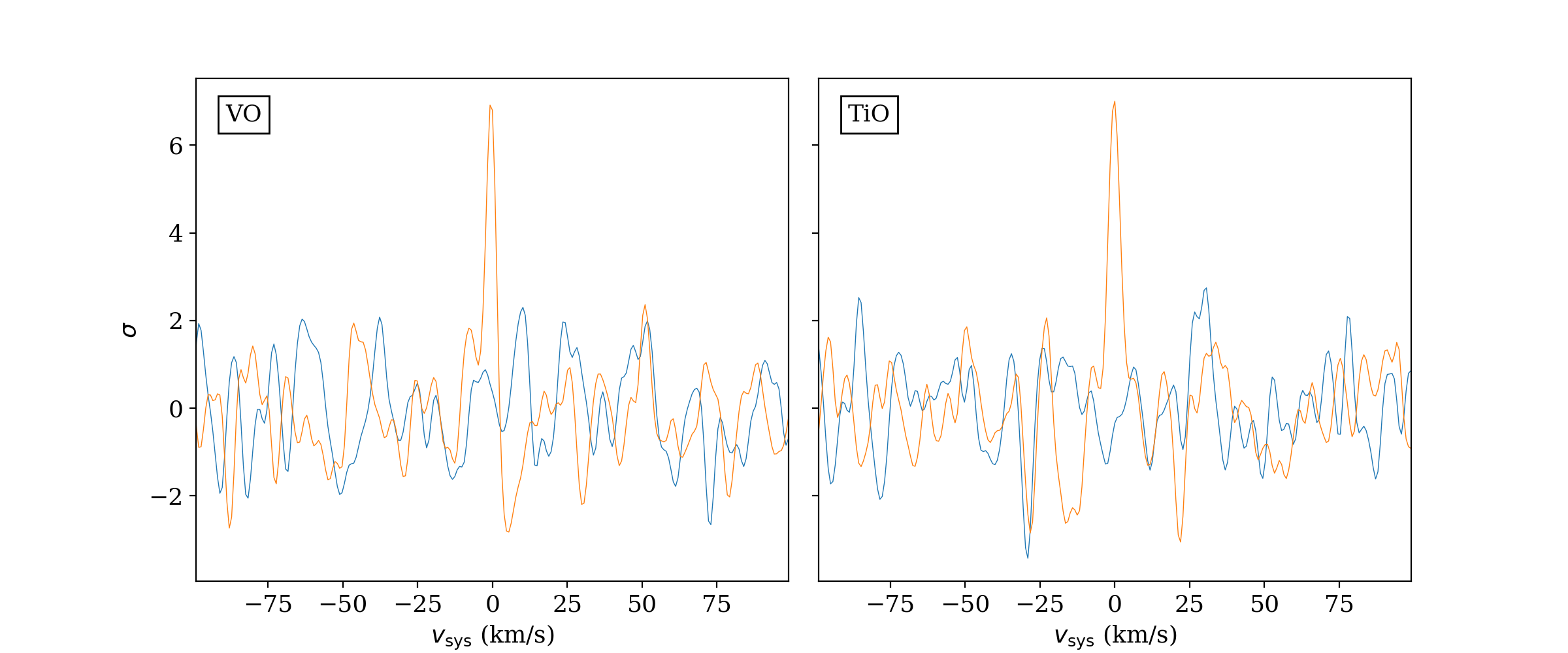}
    \caption{Slices taken from the cross-correlation maps in Fig.~\ref{fig:res} at $K_\textrm{p} = \pm$217 $\text{km s}^{-1}$, with the non-detection in blue and the injection test in orange for comparison.}
    \label{fig:slice}
\end{figure*}

No significant signal was found in any of the cross-correlation maps for TiO or VO at any volume mixing ratio, temperature or line list, whether at the expected value of $K_\textrm{p}$ or otherwise. Examples are shown in the top two rows of Fig.~\ref{fig:res}.

The possibility remains that this non-detection could be due to the predicted signal being fainter than our ability to retrieve it. To test this hypothesis, we performed a series of injection tests. The model spectra used to attempt to retrieve the signal described in Sec.~\ref{sub:model} were convolved to the instrumental resolution and multiplied into the original extracted spectra before the pre-processing steps outlined in Sec.~\ref{sub:prepro} at a $K_\textrm{p}$ of -217 $\text{km s}^{-1}$, a value chosen to have an identical velocity curve to the expected signal but reversed, to ensure that the injected signal would not be enhanced by any extant, undetected VO or TiO signal. Two of the resultant cross-correlation maps are shown in the bottom row of Fig.~\ref{fig:res}, for the \citet{Evans2018} log VMR [VO] value of -6.6 and a similar log VMR of -7 for TiO, using the 1500 K temperature also inferred by \citet{Evans2018}.  The injected signal was recovered at 6.9${\sigma}$ for VO and 7.2${\sigma}$ for TiO, demonstrating that were VO present at the abundance and temperature predicted by \citet{Evans2018}, our technique would detect it. A slice through the cross-correlation maps of both the non-detections and the injection tests at the expected value of $K_\textrm{p}$ is shown in Fig.~\ref{fig:slice}.

\begin{figure*}
	\includegraphics[width=\textwidth]{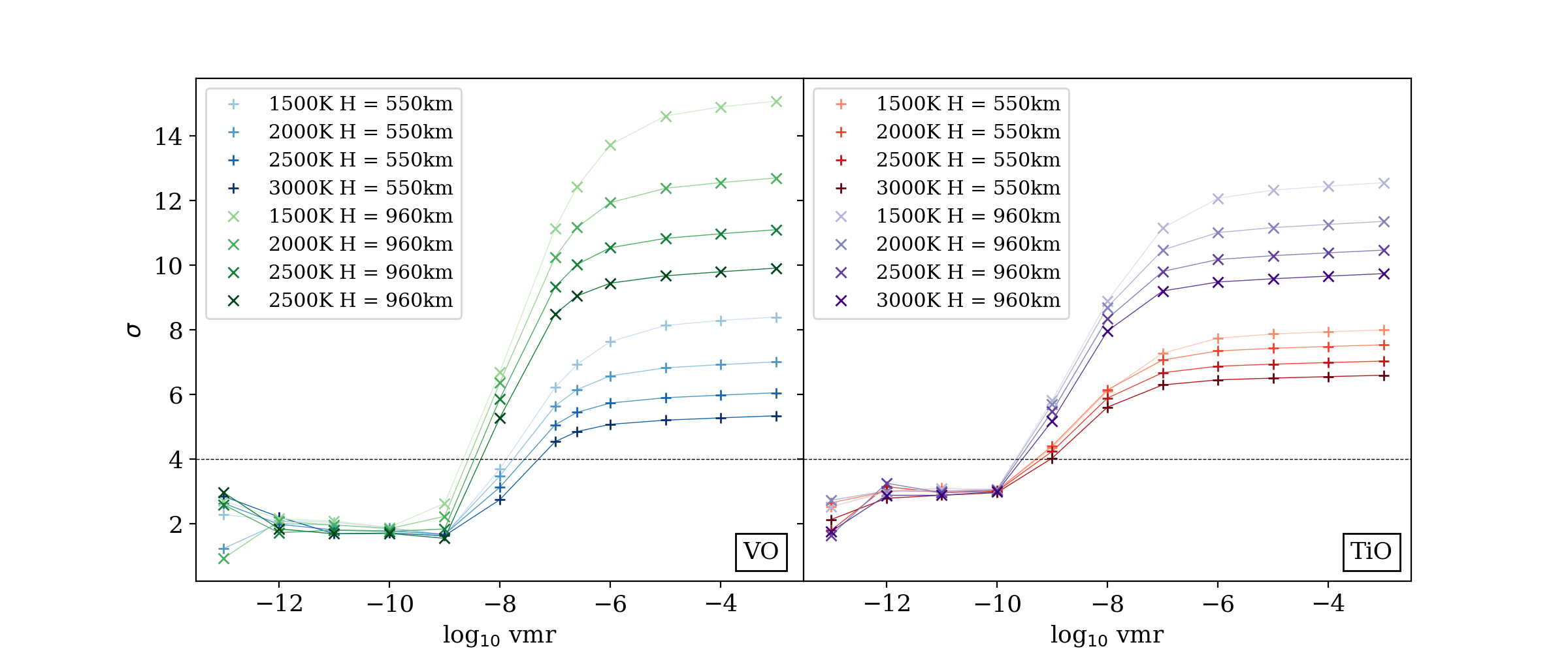}
    \caption{Detection significance vs. log volume mixing ratio abundance of VO (left) and TiO (right) from signal injection tests for all four temperatures considered. Scale height was held constant at 550km for all temperatures but 2500 K, where a scale height of 960 km was also tested.} The black dotted line indicates the chosen detection threshold of $4\sigma$.
    \label{fig:injdet}
\end{figure*}

The injection tests were also used to explore rough detection limits for VO and TiO, via the injection of templates identical in all parameters but for log VMR of the species in question. The detection threshold was once again set at 4$\sigma$, for the reasons stated in Sec.~\ref{sub:cc}. As shown in Fig.~\ref{fig:injdet}, assuming a temperature of 1500 K, by interpolating from our injection tests we can theoretically detect the presence of VO in the atmosphere of WASP-121b using our technique down to an abundance of $[\text{VO}] = -7.9$. For TiO, we can theoretically detect its presence down to an abundance of log VMR $[\text{TiO}] = -9.3$, thus placing a detection limit upon TiO lower than the limit of $[\text{TiO}] < -7.9$ placed by \citet{Evans2018}. Both of our detection limits are subsolar, where the solar abundance values are -7.1 and -8.0 for TiO and VO respectively. We also find a reduction in detection significance with temperature for both VO and TiO in our injection tests. As we held scale height constant (at 550 km) in our model spectra in order to match the previously-measured transmission spectrum of \citet{Evans2018}, the effect of increasing temperature is to change the relative line strengths of the spectral lines. As can be partially seen in Fig.~\ref{fig:model}, increasing the temperature had the effect of reducing the strength of many the strongest lines, an effect not quite offset by the strengthening of weaker lines. This led to the overall reduction in detection significance shown in Fig.~\ref{fig:injdet}. 

However, it should also be noted that 550 km is an extremely conservative estimate for the scale height of WASP-121b, based on the limb-averaged temperature of $\sim$ 1500 K presented in \citet{Evans2018}. More recent work has indicated that the limb and nightside temperatures of WASP-121b may be much higher \citep{Bourrier2019, Daylan2019, Gibson2020}, which would result in much larger scale heights: indeed, \citet{Gibson2020} retrieve a scale height of 960 km in their analysis. As a simple test of the effects of increasing scale height, we performed a linear scaling of our models to mimic the effects of increasing the scale height from 550 km to 960 km. The results, also presented in Fig.~\ref{fig:injdet}, show that increasing the scale height leads to a significant increase in detection significance and a reduction in detection limits. We choose, however, to retain the more conservative estimates gained from the lower estimate of the scale height.

A more robust method of obtaining a detection significance was also tested following \citet{Esteves2017}, in which the spectral time-series is shuffled randomly in time and the cross-correlation maps regenerated 1000 times. These maps were then used to set the pixel-by-pixel standard deviation of the original map. However, this method was found to result in similar or higher detection significance than the simpler method outlined in Sec.~\ref{sec:anal}, and we have therefore adopted the more conservative estimates from the initial method.

Caution should be applied in interpreting the results of these injection tests. Firstly, the injected signal is identical to the template used to retrieve it, which will artificially boost the detection signal, though this is not expected to do so dramatically. Secondly, there is a known degeneracy between chemical abundances and the scattering properties of the atmosphere (\citealt{Benneke2012}; \citealt{Heng2017}). For our injected spectral models, we have assumed a cloud deck at 20 mbar and a H$^{-}$ VMR of 5\,$\times$\,10$^{-10}$, consistent with \citet{Evans2018}, and our detection limits should be considered in this context. Finally, the VO line list is known to be unsuitable for high-resolution cross-correlation. We discuss this further in Sec.~\ref{sec:discus}.

\begin{figure}
	\includegraphics[width=\columnwidth]{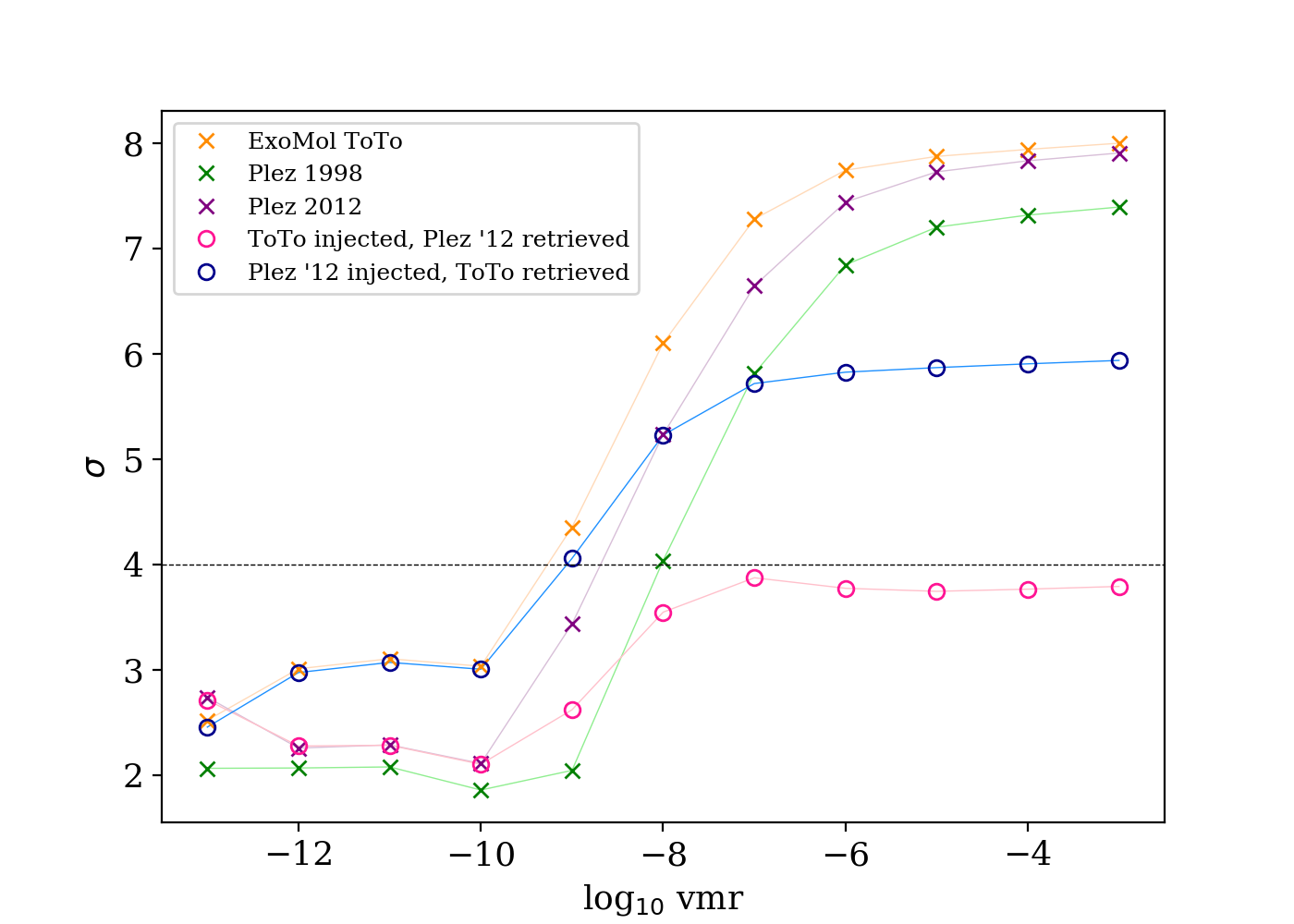}
    \caption{Detection significance vs. log volume mixing ratio abundance of TiO for injection tests using three different line lists, including the results of injecting one line list and retrieving with another. The black dotted line indicates the chosen detection threshold of $4\sigma$.}
    \label{fig:injll}
\end{figure}

We also present the results of injection tests using differing TiO line lists in Fig.~\ref{fig:injll}. While all line lists follow the same overall trend, they can be seen to behave differently. In the case of \texttt{ToTo}, the increased number of lines present in the line list is likely to be the cause of the improvement in the signal recovery. For Plez 2012, the corrections made using laboratory measurements of $^{48}$Ti$^{16}$O increase the line strength of some of the stronger lines, again boosting signal recovery.

Additionally, we investigated the behaviour of the recovered detection signal when injecting models created with one line list and retrieving with models using another. Due to the small variations in line strengths and positions between different line lists, this simple test serves as an indication of the effects of using inaccurate line lists in high-resolution cross-correlation. Further in-depth analysis of a range of TiO line lists and their effects upon signal retrieval in the cross-correlation process will be presented in Nugroho et al. \textit{(in prep)}.

The results are presented in Fig.~\ref{fig:injll}. Injecting the \texttt{ToTo} model and retrieving with Plez 2012 results in values for the detection significance which do not rise above our chosen threshold of $4\sigma$ at any log VMR: the differences between the two line lists have caused the injected signal to become essentially undetectable. However, when performing the inverse of this test, injecting models using the Plez 12 line list and retrieving using \texttt{ToTo}, the detection significance rises once again. Though it does not reach the $\sim$ $7\sigma$ level of tests using the same models for both injection and retrieval, it still results in detectable signals for a similar range of log VMRs.

The underlying cause for this difference may be due to the greater number of lines in the \texttt{ToTo} line list. The densely-overlapping lines form a pseudo-continuum in the models, which weakens overall line contrast. When cross-correlating using \texttt{ToTo} models, the greater number of lines mitigates this effect: when attempting retrieval using Plez 12, which has less lines overall, the effect is enhanced, and the signal is no longer well-recovered. This test emphasises the importance of understanding the effects inaccurate line lists can have on signal retrieval using the cross-correlation methodology: we discuss this further in Sec.~\ref{sec:discus}.

\section{Discussion}
\label{sec:discus}
Our results show non-detections of TiO and VO in the atmosphere of WASP-121b using the standard cross-correlation methodology. Additionally, our injection tests show that if VO were present at the super-solar abundances predicted in \citet{Evans2018} or \citet{Mikal-Evans2019}, we would detect the subsequent molecular signal at a high significance. Doubt had already been placed upon the existence of TiO in the limb of WASP-121b by \citet{Evans2018}, who fit their transmission spectra best with a model containing VO only; the emission spectra presented in \citet{Evans2017} and \citet{Mikal-Evans2019} is similarly best fit by models containing only VO. Our non-detection of TiO is thus consistent with previous results and adds further evidence to the non-existence of TiO in detectable quantities in the atmosphere of WASP-121b.

Conversely, our non-detection of VO contradicts results presented in \citet{Evans2016}, \citet{Evans2017}, \citet{Evans2018} and \citet{Mikal-Evans2019}, who find evidence for VO in both emission and transmission at low-resolution. It is known that low-resolution spectroscopy is troubled by the imperfect removal of systematics, especially shown in the case of STIS on HST \citep{Gibson2017, Gibson2019}, the instrument used for the transmission spectrum presented in \citet{Evans2018}. The effects of systematics on low-resolution spectra can also be seen in the improved retrieval analysis of the emission spectrum presented in \citet{Mikal-Evans2019} using WFC3 on HST. Here, they no longer fit a strong feature at 1.2 micron attributed to VO in \citet{Evans2017}: the 1.2 micron feature is now thus suspected to be an imperfectly-removed systematic. However, although in all cases their evidence for VO is inferred from best-fit models rather than a direct detection, models containing VO are best-fit in both transmission and emission, using two separate instruments. In the face of this strong evidence for the presence of VO, our non-detection seems puzzling.

However, a major question remains about the suitability of the VO line list for high-resolution spectroscopic techniques in the detection of exoplanet atmospheres. The release of the \textsc{ExoMol} group's \textsc{ToTo} line list for TiO earlier this year \citep{McKemmish2019} provided the exoplanet community with a TiO line list accurate enough at high temperatures to be used with the cross-correlation technique, after previous attempts showed that many of the current line-lists were insufficient for this task \citep{Hoeijmakers2015}. Their \textsc{\textsc{MARVEL}} codebase uses laboratory measurements to correct and refine the quantum chemistry calculations used to generate the line-lists, theoretically resulting in more accurate results \citep{McKemmish2017}. The "\textsc{MARVEL}-ised" \textsc{ToTo} line list has been found to perform slightly better than other line lists in cross-correlation with M-dwarf stars \citep{McKemmish2019}. However, the VO line list has not yet been similarly corrected. As a result, the \textsc{ExoMol} team caution that the VO line list used in this paper is probably not accurate enough for high-resolution spectroscopic exploration of exoplanet atmospheres \citetext{Sergey Yurchenko, \textit{priv. comm.}}. Until a more accurate high-temperature line-list for VO is calculated, it is impossible to state whether our non-detection is due to the true absence of gaseous VO in the limb of WASP-121b or due to an inaccurate line list weakening the strength of the recovered signal. Indeed, our own injection tests presented in Sec.~\ref{sec:res}, in which we inject models created with one line list and attempt retrieval using another in order to mimic the effects of using inaccurate line lists, show drastic changes in signal detection and significance depending on which line lists are used. Our results concerning VO should thus be interpreted in this context.

We also performed a number of injection tests in order to set a rough detection limit on the presence of VO and TiO, in which we ruled out the presence of VO and TiO at abundances greater than -7.88 and -8.18 log VMR respectively at a temperature of 1500 K. However, these figures all assume a cloud deck placed at a constant pressure of 20 mbar and  a H$^{-}$ VMR of 5\,$\times$\,10$^{-10}$. There is a well-known degeneracy between abundance and scattering properties, where the presence of clouds and hazes can mute spectral features such that a given spectral signal can be produced by either a low fractional abundance of the species in a clear atmosphere or a higher fractional abundance in the presence of clouds and hazes (\citealt{Benneke2012}; \citealt{Heng2017}). Since our method removes both the continuum and the scattering profile from our spectra, we are unable to make any measurements which could break the degeneracy. This indicates that we cannot place too much weight upon these detection limits as absolute figures but rather that they should be examined in the context of the strength of the lines appearing above the continuum. Our upper limits for both TiO and VO correspond to a line strength for the strongest lines in the template of $\sim 0.014~\Delta F$.

If we momentarily set aside concerns regarding the accuracy of the VO line list, the non-detection of TiO and VO could be caused by several factors. First, it should be noted that our spectral models (see Sec.~\ref{sub:model}) assume a well-mixed atmosphere. Atmospheric retrievals (e.g. \citealt{Daylan2019}; \citealt{Mikal-Evans2019}) show that the abundances of TiO and VO are expected to vary greatly with altitude. In the case of a very hot thermosphere ( $>$ 3000 K), TiO and VO would be dissociated at high altitude, causing a truncation of the lines present in the transmission spectrum over the altitude range probed by high-resolution spectroscopy. This could potentially cause a dilution in signal that could lead to a non-detection. Further work could focus on the injection of more sophisticated spectral models where molecular abundance varies with pressure to test whether high-altitude dissociation is strong enough to remove the possibility of detection altogether.

It has been suggested that gravitational settling could drag TiO and VO down from the upper atmosphere to the colder layer in the deeper atmosphere, where it subsequently condenses. Alternatively, the cooler temperatures on the night-side could cause TiO and VO to condense out (primarily into Ti$_3$O$_5$ and V$_2$O$_3$), and due to the high-speed winds from day- to night-side expected in tidally-locked hot Jupiters, these condensed molecules are not efficiently redistributed to the hot day-side. Thus, Ti and V remain locked in condensed form (the cold-trapping effect: \citealt{Hubeny2003}; \citealt{Spiegel2009}; \citealt{Parmentier2013}, \citeyear{Parmentier2016}; \citealt{Beatty2017}). However, the detection of Fe and Mg (\citealt{Sing2019}; \citealt{Gibson2020}; \citealt{Bourrier2020}; \citealt{Cabot2020}) in the atmosphere of WASP-121b suggests that cold-trapping does not play a large role, as we would expect Fe and Mg to be cold-trapped as well. 
Additionally, calculations from TESS phase curve measurements made by \citet{Bourrier2019} and \citet{Daylan2019} indicate a nightside temperature for WASP-121b of 2190 K, higher than the condensation temperatures of TiO and VO, although their observations probe high in the atmosphere (1 - 10 mbar) and do not rule out the possibility of cold-trapping at lower altitudes.

\citet{Parmentier2018} predicted that TiO and VO may, in the case of WASP-121b, be thermally dissociated in the dayside atmosphere; the detection of these molecules in the terminator during transmission would thus depend on their recombination timescales compared to the atmospheric recirculation timescale. However, recent work using the blue arm of the data presented in this paper by \citet{Gibson2020} retrieves a high limb temperature of $\gtrsim$ 3000 K, albeit using simplified isothermal models. When considered in the context of the atmospheric retrievals of \citet{Daylan2019}, this temperature would cause the dissociation of TiO and VO in the limb even at the lower altitudes probed by our data, reducing the log VMR of both species to $<$ -10, well below the detection limits we set with our injection tests.

It is also interesting to note that the work of \citet{Evans2017}, \citet{Evans2018} and \citet{Mikal-Evans2019} infers the presence of VO, but not TiO, when both were predicted as inversion-drivers by \citet{Hubeny2003} and \citet{Fortney2008} and were largely expected to occur together. It has been shown in the models of \citet{Goyal2019} that greater abundance of VO over TiO is only expected in a very narrow temperature range: $\sim$ 1700 - 1800 K assuming rainout condensation, or $\sim$ 1200 - 1400 K assuming local condensation. This highlights the importance of robust detections and non-detections of TiO and VO as a diagnostic for temperature and condensation processes in the atmospheres of ultra-hot Jupiters.

Despite our non-detection of TiO and VO, \citet{Evans2018} still show a feature in their transmission spectra at 600-800 nm consistent with the presence of VO (or perhaps another molecule absorbing strongly in this range). Additionally, their emission spectra (\citealt{Evans2017}; \citealt{Mikal-Evans2019}) are still best-fit with models containing VO, and clearly show water emission features consistent with the presence of a temperature inversion. It was theorised by \citet{Spiegel2009} that a hot Jupiter with a solar C/O ratio would require TiO of at least solar abundance to cause a temperature inversion, a level which both we and \citet{Evans2018} rule out in our analysis. Our analysis in this paper now seems to rule out the presence of VO also, although again we emphasise that VO cannot be robustly ruled out using high-resolution cross-correlation methodology until a more suitably accurate line list is published. Nevertheless, although TiO and VO were long thought to be the predominant cause of temperature inversions in hot Jupiters, it may now be time to look for other culprits.

Recent theoretical work has postulated a range of potential optical absorbers such as SiO, Fe, Mg, AlO, CaO and metal hydrides (\citealt{Lothringer2018}; \citealt{LothringerBarman2019}; \citealt{GandhiMadhusudhan2019}), some of which (such as AlO) have absorption bands in the 600-800 nm range. \citet{GandhiMadhusudhan2019} also suggested a theoretical link between the C/O ratio and the presence of a temperature inversion, postulating that a reduced infrared opacity due to a low H$_2$O abundance in WASP-121b could contribute towards the existence of an inversion layer; the presence of H$_2$O emission features, however, would appear to rule out this possibility. Finally, the transmission spectrum presented by \citet{Evans2018} shows a strong upward trend in the blue optical much steeper than would be consistent with Rayleigh scattering, which they tentatively identify as due to the presence of SH, another molecule known to absorb strongly in the optical and thus theoretically contribute to an inversion layer \citep{Zahnle2009}. However, this slope could also be attributed to a forest of atomic metal lines. Indeed, a recent analysis by our group of the blue arm of the UVES data used in this paper \citep{Gibson2020} found strong evidence for the presence of \ion{Fe}{i} in the atmosphere of WASP-121b, a discovery supported by recent analyses of HARPS high-resolution spectra by \citet{Bourrier2020} and \citet{Cabot2020}. \ion{Fe}{i} has strong absorption at blue-optical wavelengths, and as WASP-121 is an F6-type star, this could lead to significant amounts of energy being deposited at high altitudes and could potentially be the sole or predominant cause of the observed temperature inversion. However, \ion{Fe}{i} cannot be responsible for the absorption feature seen in the transmission spectra at 600-800 nm in the transmission spectrum of \citet{Evans2018}, and if not caused by VO, the presence of another optical absorber with strong absorption in this range cannot be ruled out. 

Future searches could focus on the detection of a molecular species with strong absorption features in this range, such as AlO and MgH, in order to constrain the origin of this feature. Indeed, a tentative detection of AlO has already been reported in the atmosphere of ultra-hot WASP-33b \citep{vonEssen2019}, indicating that this may be a productive avenue of inquiry. Alternatively, should a more robust line list for VO be produced, a reanalysis of the data presented in this work could more definitively confirm or rule out the presence of VO in WASP-121b.

\section{Conclusions}
\label{sec:con}
We have presented a non-detection of TiO and VO in the atmosphere of the ultra-hot Jupiter WASP-121b, a planet previously shown to have evidence for a temperature inversion and for the existence of VO (but not necessarily TiO). Using injection tests, we have shown that were VO present in the abundance predicted by \citet{Evans2018}, we could retrieve the molecular signal using our technique with relative ease. Additionally, we have placed rough constraints on the presence of VO in the limb of WASP-121b using injection tests, and find that -- while bearing in mind the established degeneracy between scattering properties and abundance -- we can detect the presence of VO down to a sub-solar log volume mixing ratio of -7.88, and TiO down to a sub-solar log volume mixing ratio of -9.26. The absence of TiO and VO in transmission could be due to thermal dissociation of these molecules on the hot dayside of WASP-121b in conjunction with a slow recombination timescale, or due to cold-trapping locking condensed TiO and VO on the colder nightside. However, we also emphasise that the VO line-list utilised for this paper is thought to lack the accuracy in line position required for high-resolution spectroscopic searches for molecular species. Despite the strength of our detection in our injection tests, it is nevertheless clear that a definitive answer on the existence of VO in the limb of WASP-121b using high-resolution spectroscopy will be dependent on a more accurate, updated line-list for VO. 

Recent work from our team \citep{Gibson2020} posits a different source species (\ion{Fe}{i}) for the observed temperature inversion in WASP-121b: however, \ion{Fe}{i} cannot be responsible for the observed absorption feature at 600-800 nm in the transmission spectrum presented by \citet{Evans2018}. If physical, and if not caused by the presence of VO, a range of molecular species could be responsible for this feature, and future work will focus on the search for these and for many other such molecules theorised to be present in the atmospheres of ultra-hot Jupiters.

\begin{acknowledgements}
The authors are very grateful to the anonymous referee for their insightful comments and suggestions. This work is based on observations collected at the European Organisation for Astronomical Research in the Southern Hemisphere under ESO programme 098.C-0547. S.R.M. acknowledges funding from the Northern Ireland Department for the Economy. N.P.G. gratefully acknowledges support from Science Foundation Ireland and the Royal Society in the form of a University Research Fellowship. S.K.N. and C.A.W. would like to acknowledge support from UK Science Technology and Facility Council grant ST/P000312/1. We are exceptionally grateful to the developers of the \textsc{NumPy}, \textsc{SciPy}, \textsc{Matplotlib}, \textsc{iPython}, \textsc{scikit-learn} and \textsc{AstroPy} packages, which were used extensively in this work (\citealt{vanderWalt2011}; \citealt{Virtanen2019}; \citealt{Hunter2007}; \citealt{PerezGranger2007}; \citealt{Pedregosa2012}; \citealt{Astropy2013}). The perceptually-uniform scientific colour map used in Figs.~\ref{fig:meth} and~\ref{fig:res} is Acton by \citet{Crameri2018}.
\end{acknowledgements}

%
%

\bibliographystyle{aa} 
\bibliography{W121b-TioVO_bibs} 

\end{document}